\documentclass[aps,twocolumn,showpacs]{revtex4}
\def\comment#1{}

\def\lfrac#1#2{{#1/#2}}

\begin{document}

%\preprint{}

\title{Effective Lagrangian of QED}

%\author{Remo Ruffini$^{(a)}$, She-Sheng Xue$^{(a)}$, and
%Hagen Kleinert$^{(b)}$}

\author{
Remo Ruffini and She-Sheng Xue
}

%\email{kleinert@physik.fu-berlin.de  ; ruffini@icra.it ; xue@icra.it}

\affiliation{
%(a)
ICRANeT and
Physics Department, University of Rome ``La Sapienza", 00185 Rome, Italy\\
%(b) Institute of Theoretical Physics, Free University of Berlin, Arnimallee 14, D-14195,
%Berlin, Germany
}
%\date{August, 2000}

%\date{\today}

\begin{abstract}

From the Euler-Heisenberg formula
we calculate the exact real part of the
one-loop effective Lagrangian
of Quantum Electrodynamics
in a constant electromagnetic field, and determine
its strong-field limit.

\end{abstract}


\maketitle

\subsubsection{Introduction}\label{introduction}

The possibility of electron-positron pair production from the vacuum of Quantum Electrodynamics (QED)
vacuum was first pointed out by Sauter, Euler, Heisenberg and Schwinger \cite{sauter,euler,schwinger}
who studied
the behavior of the
Dirac vacuum in a strong external electric field.
If the field is stronger than the critical value $E_c=m^2c^3/e\hbar$,
the energy of the vacuum can be lowered by spontaneously
creating an electron-positron pair.
This is the {\em Sauter-Euler-Heisenberg-Schwinger process\/}.
For many decades, both theorists and experimentalists have been interested in the aspects of
electron-positron pair production from the QED vacuum by an external electromagnetic field.
There are many reasons for the interest in the phenomenon of pair-production in a strong electric field.
In addition to understand the behavior of QED in the strong coupling regime,
the most compelling one is that both laboratory conditions and astrophysical events may
now exist for observing this process.

The above process is governed by the imaginary part of the
Euler-Heisenberg Lagrangian, and it is this imaginary
part which has received most attention until now.
In this article, we supplement the
existing results by
exact expressions for
the real part.
This determines
the lowering of the energy of the vacuum
by the external fields.

\subsubsection{QED in external electromagnetic fields}

The QED Lagrangian describing the interacting system
of photons, electrons and positrons reads
\begin{equation}
{\mathcal L} = {\mathcal L}_0^\gamma + {\mathcal L}^{e^+e^-}_0 + {\mathcal L}_{\rm int},
\label{qcdl}
\end{equation}
where the
free Lagrangians
${\mathcal L}^{e^+e^-}_0$ and
${\mathcal L}_0^\gamma$ for electrons and photons
are expressed
in terms of Dirac field
$\psi(x)$ and an
electromagnetic field $A_\mu(x)$
 as follows:
\begin{eqnarray}
{\mathcal L}^{e^+e^-}_0 &=&~\, \bar\psi(x)(i\gamma^\mu\partial_\mu
-m_e)\psi(x),\label{L0pair} \\
{\mathcal L}_0^\gamma ~~&=& -{1\over4}F_{\mu\nu}(x)F^{\mu\nu}(x)+
{\rm gauge\!-\!fixing~term}.\label{L0gamma}
\end{eqnarray}
Here $ \gamma ^\mu$ are the $4\times4$ Dirac matrices,
$\bar \psi(x)\equiv \psi^\dagger(x) \gamma ^0$,
and
 $F_{\mu \nu }=\partial_\mu A_\nu-\partial_\nu A_\mu$
denotes the electromagnetic  field
tensor.
Minimal coupling gives rise to the
interaction Lagrangian
\begin{eqnarray}
{\mathcal L}_{\rm int} &=& -e\bar\psi(x)\gamma^\mu \psi(x) A_\mu(x).
\label{L0int}
\end{eqnarray}
We are using natural units $\hbar = c=1$.

An external electromagnetic field
 is incorporated
by adding to the quantum field
$A_\mu$ in (\ref{L0int})
an unquantized
external vector potential $A^{\rm e}_\mu$, so that the total
 interaction becomes
\begin{equation}
{\mathcal L}_{\rm int}+
{\mathcal L}^{\rm e}_{\rm int}
 = -e\bar\psi(x)\gamma^\mu
\psi(x)
\left[
A_\mu(x)+
A^{\rm e}_\mu(x)\right] .
\label{intc}
\end{equation}
Quantum field theory is defined by
a functional integral formulation for the quantum mechanical partition function
\begin{equation}
Z[A^{\rm e}]=\int [{\mathcal D}\psi {\mathcal D}\bar\psi {\mathcal D}A_\mu]
\exp \left[ i\int d^4x ({\mathcal L} + {\mathcal L}^{\rm e}_{\rm int}
) \right],
\label{path}
\end{equation}
to be integrated over all
fluctuating electromagnetic and Grassmannian electron fields.
The normalized quantity $Z[A^{\rm e}]$ gives the
amplitude
 for the vacuum to vacuum transition in the presence of the external
classical electromagnetic field:
\begin{equation}
\langle {\rm out}, 0|0, {\rm in}\rangle = {Z[A^{\rm e}]\over Z[0]},
\label{vvamplitude}
\end{equation}
where $|0, {\rm in}\rangle$ is the initial vacuum state at the time
$t=t_-\rightarrow -\infty$, and
$\langle {\rm out}, 0|$  the final vacuum state at the time $t=t_+\rightarrow +\infty$.
By selecting only the one-particle irreducible
Feynman diagrams
in the the perturbation expansion of $Z[A^{\rm e}]$
one obtains the effective action as a functional of $A^{\rm e}$
\begin{equation}
\Delta{\mathcal A}_{\rm eff}[A^{\rm e}]\equiv -i\ln \langle {\rm out}, 0|0, {\rm in}\rangle.
\label{eaction}
\end{equation}
In general, there exists no local effective Lagrangian density
$\Delta{\mathcal L}_{\rm eff}$ whose spacetime
integral is  $\Delta{\mathcal A}_{\rm eff}[A^{\rm e}]$. An infinite set of derivatives
would be needed, i.e., $\Delta{\mathcal L}_{\rm eff}$ would have the arguments
$A^{\rm e}(x),\partial_\mu A^{\rm e}(x), \partial_\mu\partial _\nu A^{\rm e}(x), \dots~$, containing
gradients of arbitrarily high order.
With presently available methods it is possible to calculate a few terms in such
a gradient expansion, or a semi classical approximation {\it \`a la} WKB or an arbitrary but
smooth spacetime dependence.
Under the assumption that the external field $A^{\rm e}(x)$ varies smoothly over a finite region
spacetime, we may define an approximately local effective Lagrangian $\Delta{\mathcal L}_{\rm eff}[A^{\rm e}(x)]$,
\begin{equation}
\Delta{\mathcal A}_{\rm eff}[A^{\rm e}]\simeq \int d^4x \Delta{\mathcal L}_{\rm eff}[A^{\rm e}(x)]
\approx V\Delta t \Delta{\mathcal L}_{\rm eff}[A^{\rm e}],
\label{effl}
\end{equation}
where $V$ is the spatial volume and time interval $\Delta t=t_+-t_-$.

For a large time interval $\Delta t=t_+-t_-\rightarrow \infty$, the amplitude of the vacuum
to vacuum transition (\ref{vvamplitude}) has the form,
\begin{equation}
\langle {\rm out}, 0|0 ,{\rm in}\rangle = e^{-i(\Delta{\mathcal E}_0-i\Gamma/2)\Delta t},
\label{vvamplitude1}
\end{equation}
where $\Delta{\mathcal E}_0={\mathcal E}_0(A^{\rm e})-{\mathcal E}_0(0)$ is
the difference between the vacuum energies in the presence and the absence of
the external field, $\Gamma$ is the vacuum decay rate, and $\Delta t$ the time over which
field is non zero.
The probability that the vacuum remains as it is in the presence of the external
classical electromagnetic field is
\begin{equation}
|\langle {\rm out}, 0|0 ,{\rm in}\rangle|^2 = e^{-2{\rm Im}\Delta{\mathcal A}_{\rm eff}[A^{\rm e}]}.
\label{probability}
\end{equation}
This determines the decay rate of the vacuum in an external electromagnetic field:
\begin{equation}
\frac{ \Gamma}{V}= \frac{2{\rm \,Im}\Delta{\mathcal A}_{\rm eff}[A^{\rm e}]}{\Delta t V}
\approx 2{\rm \,Im}\Delta{\mathcal L}_{\rm eff}[A^{\rm e}].
\label{path21}
\end{equation}
The finite lifetime is caused by the production of electron and positron pairs.
The external field changes the energy density by
\begin{equation}
\frac{ \Delta{\mathcal E}_0}{V}= -\frac{{\rm \,Re}\Delta{\mathcal A}_{\rm eff}[A^{\rm e}]}{\Delta t V}
\approx -{\rm \,Re}\Delta{\mathcal L}_{\rm eff}[A^{\rm e}].
\label{path21e}
\end{equation}

\subsubsection{Schwinger formula for pair production}\label{Schwingerformula}

The Dirac field appears quandratically in the
partition functional (\ref{path}) and can be integrated out,
leading to
\begin{eqnarray}
Z[A^{\rm e}] &=&\int  {\mathcal D}A_\mu\,
{\rm Det}
\{i\!\!\not{\!\partial}-e[\not{\hspace{-5pt}A(x)}+
\not{\hspace{-5pt}A}^{\rm e}(x)]
-m_e+i\eta\};\nonumber\\
\not{\!\partial}& \equiv &\gamma^\mu\partial_\mu,
\end{eqnarray}
where Det denotes the functional determinant
of the Dirac operator.
Ignoring the fluctuations of the electromagnetic field, the result is a functional
of the external vector potential  $A^{\rm e}(x)$:
\begin{equation}
Z[A^{\rm e}]\approx {\rm const.}\times
{\rm Det}
\{i\!\!\not{\!\partial}-e
\not{\hspace{-5pt}A}^{\rm e}(x)-m_e+i\eta\}.
\label{path1}
\end{equation}
 which is the one-loop approximation of ${\mathcal O}(\hbar)$.
The infinitesimal constant $i\eta$ with $ \eta >0$
specifies the treatment of singularities in energy integrals.
>From Eqs.~(\ref{vvamplitude})--(\ref{path1}),
the effective action (\ref{probability}) is
given by
\begin{eqnarray}
 \Delta{\mathcal A}_{\rm eff}[A^{\rm e}] &=& -i
{\rm Tr}\ln \Big\{\left[ i\!\!\not{\!\partial}-
e\not{\hspace{-5pt}A}^{\rm e}(x)-m_e+i\eta\right]\nonumber\\
&\cdot&
{1\over
 i\!\!\not{\!\partial}
-m_e+i\eta}\Big\},
\label{eaction12}
\end{eqnarray}
where Tr denotes the functional and Dirac trace. In physical unit, this is of order $\hbar$. The result may be
expressed as a one-loop Feynman diagram, so that one speaks of a one-loop approximation.
More convenient will be the equivalent
expression
\begin{eqnarray}
\Delta{\mathcal A}_{\rm eff}[A^{\rm e}] &=&
-\frac{i}{2}\,{\rm Tr}\ln \Big(\{[i\!\!\not{\!\partial}-
e\not{\hspace{-5pt}A}^{\rm e}(x)]^2-m_e^2+i\eta\}\nonumber\\
&\cdot& {1\over -\partial ^2-m_e^2+i\eta}\Big),
\label{eaction1}
\end{eqnarray}
where
\begin{equation}
[i\!\!\not{\!\partial}-
e\not{\hspace{-5pt}A}^{\rm e}(x)]^2=
[i\partial_\mu-eA^{\rm e}_\mu(x)]^2+{e\over2}\sigma^{\mu\nu}F^{\rm e}_{\mu\nu}.
\label{sqrt}
\end{equation}
Using the identity
\begin{equation}
\ln{a_2\over a_1}=\int_0^\infty {ds\over s}\big[ e^{is(a_1+i\eta)}-e^{is(a_2+i\eta)}\big],
\label{identityab}
\end{equation}
Eq.~(\ref{eaction1}) becomes the Schwinger form
\begin{eqnarray}
\Delta{\mathcal A}_{\rm eff}[A^{\rm e}]
& = &\frac{i}{2}\int_0^\infty {ds\over s}e^{-is(m_e^2-i\eta)}\label{probability01}\\
&\!\cdot\!&\!\!\!
{\rm Tr}
\langle x| e^{ is\left\{[i\partial_\mu-eA^{\rm e}_\mu(x)]^2
+ {e\over2}\sigma^{\mu\nu}F^{\rm e}_{\mu\nu}\right\}}
-e^{-is\partial ^2}|x\rangle\nonumber ,
\end{eqnarray}
where $\sigma^{\mu\nu}\equiv \frac{i}{2}[\gamma^\mu,\gamma^\nu]$,\,
$F^{\rm e}_{\mu\nu}=\partial_\mu A^{\rm e}_\nu-\partial_\nu A^{\rm e}_\mu$, and
$\langle x|\{\cdot\cdot\cdot\}|x\rangle $ are the
diagonal matrix elements
in the local basis $|x\rangle $. This is defined {\it \`a la} Dirac by the
matrix elements with the momentum eigenstates $|k\rangle$ being
 plane waves:
  $\langle x|k\rangle=
e^{-ikx}$. The symbol ${\rm Tr}$
denotes integral $\int d^4x$ in spacetime and the trace
in spinor space.
For constant electromagnetic fields, the integrand
 in  (\ref{probability01}) does
 not depend on $x$, and
$\sigma^{\mu\nu}F^{\rm e}_{\mu\nu}$ commutes with all other operators.
This will allow us
to calculate
the exponential in Eq.~(\ref{probability01}) explicitly.
The presence of $-i \eta $ in the mass term ensures the convergence of integral for $s\rightarrow\infty$.

If only a constant electric field ${\bf E}$ is present, it may
be assumed to point
along the
${\hat{\bf z}}$-axis, and one can choose a gauge such
that $A^{\rm e}_z=-Et$ is the only nonzero component of $A^{\rm e}_\mu$. Then one finds
\begin{equation}
{\rm tr}\exp is\left[{e\over2}\sigma^{\mu\nu}F^{\rm e}_{\mu\nu}\right]=4\cosh(seE),
\label{iz1}
\end{equation}
where the symbol ${\rm tr}$ denotes the trace in spinor space.
Using commutation relation $[\partial _0,x^0]=1$, where $x^0=t$,
one computes the exponential term in the effective action
(\ref{probability01}) (c.e.g. \cite{book-Itzykson})
\begin{eqnarray}
&&\langle x| \exp is\left[(i\partial _\mu-eA^{\rm e}_\mu(x))^2+{e\over2}
\sigma^{\mu\nu}F^{\rm e}_{\mu\nu}\right]
|x\rangle\nonumber\\
 &=& {eE\over (2\pi)^2is}\coth(eEs).
\label{iz2}
\end{eqnarray}
The second term in Eq.~(\ref{probability01}) is obtained by setting $E= 0$ in Eq.~(\ref{iz2}), so that
then the effective action (\ref{probability01}) yields,
\begin{eqnarray}
\Delta{\mathcal A}_{\rm eff} &=&
{1\over 2(2\pi)^2}\int d^4x \int_0^\infty {ds\over s^3}
\left[eEs\coth(eEs)-1\right]\nonumber\\
&\cdot& e^{-is(m^2_e-i\eta)}.
\label{oneloops1}
\end{eqnarray}
Since the field is constant, the
integral over $x$ gives a volume factor, and so the
effective action
(\ref{probability01}) can be attributed to the spacetime integral over an
effective Lagrangian (\ref{effl})
\begin{equation}
\Delta{\mathcal L}_{\rm eff}
=
{1\over 2(2\pi)^2}\int_0^\infty {ds\over s^3}
\left[eEs\coth(eEs)-1\right]
e^{-is(m^2_e-i\eta)}.
\label{oneloopl1us}
\end{equation}

By expanding the integrand in Eq.~(\ref{oneloopl1us}) in powers of $e$, one obtains,
\begin{eqnarray}
&&{1\over s^3}\left[eEs\coth(eEs)-1\right]
e^{-i s(m^2_e-i\eta)}\nonumber\\
&&= \left[{e^2\over 3s}E^2-{e^4s\over 45}E^4 +{\mathcal O}(e^6)\right]
e^{-i s(m^2_e-i\eta)}.
\label{oneloopl1s20}
\end{eqnarray}
The small-$s$ divergence in the integrand,
\begin{equation}
{e^2\over 3}E^2
{1\over 2(2\pi)^2}\int_0^\infty {ds\over s}e^{-is(m^2_e-i\eta)},
\label{oneloopl1s2}
\end{equation}
is proportional to the electric field in the original Maxwell Lagrangian.
The divergent term (\ref{oneloopl1s2}) can therefore be removed by a renormalization
of the field $E$.
Thus, we add subtract counterterm in Eq.~(\ref{oneloopl1us})and form,
\begin{eqnarray}
\Delta{\mathcal L}_{\rm eff}
&=&
{1\over 2(2\pi)^2}\int_0^\infty {ds\over s^3}
\left[eEs\coth(eEs)-1-{e^2\over 3}E^2s^2\right]\nonumber\\
&\cdot& e^{-is(m^2_e-i\eta)}.
\label{oneloopl1}
\end{eqnarray}
Remembering Eq.~(\ref{path21}), we find
from (\ref{oneloopl1})  the decay rate of the vacuum per unit volume
\begin{eqnarray}
\frac{ \Gamma }{V}
&=&{1\over (2\pi)^2}{\rm Im}\int_0^\infty {ds\over s^3}\left[eEs\coth(eEs)-1-{e^2\over 3}E^2s^2\right]\nonumber\\
&\cdot& e^{-is(m_e^2-i\eta)}.
\label{gprobability}
\end{eqnarray}

The integral (\ref{gprobability}) can be evaluated analytically by
proceeds by the
method of residues.
Since the integrand is even,
the integral can be extended
to the entire $s$-axis. After this, the
integration contour is deformed to enclose the
negative imaginary axis
and to pick up
the contributions of the poles of the $\coth$ function
at $s=n\pi/eE$.
The result is
\begin{equation} \!\!\!\!\!\!\!\!
\frac{ \Gamma }{V}
={\alpha E^2\over \pi^2}\sum_{n=1}^\infty {1\over n^2}\exp
\left(-{n\pi E_c\over E}\right),
\label{probability1}
\end{equation}
where $E_c=m^2/e$ is the value of critical field. This result due to  Schwinger \cite{schwinger}
is valid to lowest order in $\hbar$
for arbitrary constant electric field strength.

An analogous calculation for a charged scalar field yields
\begin{equation}
\frac{ \Gamma }{V}
={\alpha E^2\over 2\pi^2}\sum_{n=1}^\infty {(-1)^{n+1}\over n^2}\exp
\left(-{n\pi E_c\over E}\right).
\label{bosonrate}
\end{equation}
These Schwinger results complete
the development.
The leading $n=1$ -terms agree with
the WKB results obtained by Sauter \cite{sauter} and Heisenberg-Euler \cite{euler}.

\subsubsection{Pair production in constant electromagnetic fields}\label{EandB}

If the constant external field has both ${\bf E}$ and ${\bf B}$ nonzero, the effective action will
be a function of the two Lorentz invariants. We can now go to an arbitrary
Lorentz frame by expressing the result in terms of the scalar $S$ and the pseudoscalar $P$,
that can be formed
from arbitrary electromagnetic fields
${\bf B}$ and ${\bf E}$ fields:
\begin{equation}    \!\!\!\!\!\!
 S \equiv
\frac{1}{4}F_{\mu \nu }F^{\mu \nu }=
\frac{1}{2}({\bf E}^2-{\bf B}^2);
\quad
  P\equiv
\frac{1}{4}F_{\mu \nu }\tilde F^{\mu \nu }=
 {\bf E}\cdot {\bf B},
\label{lightlike}
\end{equation}
where
$
\tilde F^{\mu \nu }\equiv  \epsilon ^{\mu \nu  \lambda  \kappa }F_{ \lambda  \kappa }$ is the dual field tensor.
It is useful to define the related invariants
$ \varepsilon $ and $ \beta $
as the solutions of the invariant equations
\begin{equation}
\varepsilon ^2- \beta ^2\equiv {\bf E}^2-{\bf B}^2\equiv 2 S,\quad
 \varepsilon \beta \equiv {\bf E}\cdot{\bf B}\equiv P,
\label{ab}
\end{equation}
which read explicitly
\begin{eqnarray} 
\left\{
{ \varepsilon \atop
 \beta }\right\}
 & \equiv\!&
\frac{1}{ \sqrt{2} }
 \sqrt{
 \sqrt{
 ({\bf E}^2-{\bf B}^2)^2+4( {\bf E}\cdot {\bf B})^2
}
\pm ({{\bf E}^2-{\bf B}^2})
}
%&=&\!\frac{1}{ \sqrt{2} }
% \sqrt{
% \sqrt{ S^2+P^2
%}
%\pm S
%}
\nonumber\\
&=&\sqrt{(S^2+P^2)^{1/2}\pm S}.
\label{fieldinvariant}
\end{eqnarray}
Then exists a special Lorentz frame to be called the {\it center-of-fields frame}, where
the fields ${\bf B}_{\rm CF}$ and ${\bf E}_{\rm CF}$ are parallel. In this frame,
$ \beta  =|{\bf B}_{\rm CF}|$ and $ \varepsilon =|{\bf E}_{\rm CF}|$.

Relations (\ref{ab}) are invariant under the discrete duality transformation:
\begin{equation}
|{\bf B}|\rightarrow -i|{\bf E}|,\quad |{\bf E}|\rightarrow i|{\bf B}|,
\label{duality1}
\end{equation}
i.e., under
\begin{equation}
\beta\rightarrow -i\varepsilon ,\quad \varepsilon\rightarrow i\beta.
\label{duality2}
\end{equation}
This implies that effective action for a pure magnetic field can be simply obtained
by replacing $E|\rightarrow iB$
in Eqs.~(\ref{iz2},\ref{oneloopl1}), so that
\begin{eqnarray}
&&\langle x| \exp is\left[(i\partial _\mu-eA^{\rm e}_\mu(x))^2+{e\over2}\sigma^{\mu\nu}F^{\rm e}_{\mu\nu}\right]
|x\rangle \nonumber\\
&&={eB\over (2\pi)^2is}\cot(eBs),
\label{iz2b}
\end{eqnarray}
and
\begin{eqnarray}
\Delta{\mathcal L}_{\rm eff}
&=&
{1\over 2(2\pi)^2}\int_0^\infty {ds\over s^3}
\left[eBs\cot(eBs)-1+{e^2\over 3}B^2s^2\right]\nonumber\\
&\cdot& e^{-is(m^2_e-i\eta)}.
\label{oneloopl1b}
\end{eqnarray}

If both electric and magnetic fields ${\bf E}$ and ${\bf B}$ are present,
we assume the parallel ${\bf E}_{\rm CF}$ and ${\bf B}_{\rm CF}$ fields point
along the ${\hat{\bf z}}$-axis in {\it center-of-fields frame}.
We further choose a gauge such
that only $A^{\rm e}_z=-E_{\rm CF}t$, $A^{\rm e}_y=B_{\rm CF}x^1$ are nonzero.
The exponential in the effective action Eq.~(\ref{probability01})
can then be factorized into a product of the magnetic part and the electric part.
Based on the duality, we obtain directly
\comment{ Armed with the same method used to compute the electric part (\ref{iz1},\ref{iz2}), one
can compute the magnetic part by using the communication relation
$[\partial _1,x^1]=1$, where $x^1=x$. Or one can make the substitution
(\ref{duality1}) to obtain the magnetic part, based on the discrete symmetry
of duality. As results, Eqs.~(\ref{iz1},\ref{iz2}) become }
\begin{equation}
{\rm tr}\exp is\left[{e\over2}\sigma^{\mu\nu}F^{\rm e}_{\mu\nu}\right]
=4\cosh(seE_{\rm CF})\cos(seB_{\rm CF}),
\label{iz1eb}
\end{equation}
and
\begin{eqnarray}
&\langle x| \exp is\left\{[i\partial _\mu-eA^{\rm e}_\mu(x)]^2+{e\over2}\sigma^{\mu\nu}F^{\rm e}_{\mu\nu}\right\}
|x\rangle \nonumber\\
&=\displaystyle\frac{1}{(2\pi)^2}{eE_{\rm CF}\over is}\coth(seE_{\rm CF})
{eB_{\rm CF}\over s}\cot(seB_{\rm CF}).
\label{iz2eb}
\end{eqnarray}
Performing the same substraction as before, we obtain the effective Lagrangian
\begin{eqnarray}
\Delta{\mathcal L}_{\rm eff} &=&\displaystyle
{1\over 2(2\pi)^2}\int_0^\infty {ds\over s^3}
\Big[e^2E_{\rm CF}B_{\rm CF} s^2\coth(seE_{\rm CF} )
\nonumber\\
&\cdot & \cot(seB_{\rm CF})
-1-{e^2\over 3}(E^2_{\rm CF}-B^2_{\rm CF})s^2
\Big]\nonumber\\
&\cdot & e^{-is(m^2_e-i\eta)}.
\label{onelooplcf}
\end{eqnarray}
In an arbitrary Lorentz frame, $E_{\rm CF}$ and $B_{\rm CF}$ are replaced
by the invariants $\varepsilon$ and $\beta$:
\begin{eqnarray}
\Delta{\mathcal L}_{\rm eff}
\!&=&\!\displaystyle
{1\over 2(2\pi)^2}\int_0^\infty {ds\over s^3}
\Big[e^2\varepsilon\beta s^2\coth(e\varepsilon s )\cot(e\beta s )\nonumber\\
&-&1-{e^2\over 3}(\varepsilon^2-\beta^2)s^2\Big]
e^{-is(m^2_e-i\eta)}.
\label{oneloopl}
\end{eqnarray}
The decay rate becomes
\begin{eqnarray}
\frac{ \Gamma }{V}& = &\displaystyle{1\over (2\pi)^2}{\rm Im}\int_0^\infty {ds\over s^3}
\Big[e^2\varepsilon\beta s^2\coth(e\varepsilon s )\cot(e\beta s )\nonumber\\
&-& 1-{e^2\over 3}(\varepsilon^2-\beta^2)s^2\Big]
e^{-is(m_e^2-i\eta)}.
\label{gprobabilityeb}
\end{eqnarray}

The integral is evaluated as in Eq.~(\ref{probability1}) by the
method of residues, and yields \cite{schwinger}
\begin{equation}
\frac{ \Gamma }{V}={  \alpha   \varepsilon^2\over \pi^2 }
\sum_{n=1}  \frac{1}{n^2}
\frac{ n\pi\beta / \varepsilon }
{\tanh {n\pi \beta/ \varepsilon}}\exp\left(-{n\pi E_c\over \varepsilon}\right),
\label{probabilityeh}
\end{equation}
which reduces for $\beta \rightarrow 0$ (${\bf B}=0$) correctly to (\ref{probability1}).

The analogous result for bosonic fields is
\begin{equation}
\frac{ \Gamma }{V}={  \alpha   \varepsilon^2\over 2\pi^2 }
\sum_{n=1}  \frac{(-1)^n}{n^2}
\frac{ n\pi\beta / \varepsilon }
{\sinh {n\pi \beta/ \varepsilon}}\exp\left(-{n\pi E_c\over \varepsilon}\right),
\label{probabilityehb}
\end{equation}
Note that the magnetic field produces in the fermionic case a extra factor
$\lfrac{( n\pi\beta / \varepsilon )}
{\tanh ({n\pi \beta/ \varepsilon}})>1$ in each term
which enhances the decay rate. The bosonic series (\ref{probabilityehb}), on the other hand,
carries in each term  a suppression factor
$\lfrac{ (n\pi\beta / \varepsilon )}
{\sinh {n\pi \beta/ \varepsilon}}<1$.

The decay rate $ \Gamma /V$
gives the
number of electron-positron pairs produced per unit volume.
The prefactor can be estimated on dimensional grounds.
It has
the dimension of $E_c^2/\hbar $, i.e.,
$m^4 c^5/\hbar^4$. This arises from the energy of a pair
$2m_ec^2$ divided by
the volume
whose diameter is the Compton wavelength
$\hbar/m_ec$ produced within a Compton time
$\hbar/m_ec^2$.
The exponential factor suppresses
pair production as long as the
electric field is much smaller
than the critical electric field $E_c$.
The general results ({\ref{probabilityeh}),(\ref{probabilityehb})
was first obtained by Schwinger
\cite{schwinger} for scalar and spinor electrodynamics.

\subsubsection{Effective nonlinear Lagrangian}\label{nonlinear}

In this section, we evaluate further the effective Lagrangian
(\ref{oneloopl}) for arbitrary constant electromagnetic fields ${\bf E}$ and ${\bf B}$.
Making the expansions,
\begin{eqnarray}
e\varepsilon s\coth(e\varepsilon s )
&=\displaystyle\sum_{n=-\infty}^{\infty}{s^2 \over(s^2+\tau^2_n)};\quad
\tau_n\equiv n\pi/e\varepsilon,\label{cosnm0}\\
e\beta s\cot(e\beta s )
&=\displaystyle\sum_{m=-\infty}^{\infty}{s^2 \over(s^2-\tau^2_m)},\quad
\tau_m\equiv m\pi/e\beta,
\label{cosnm}
\end{eqnarray}
we obtain the finite effective Lagrangian,
\begin{eqnarray}
\Delta{\mathcal L}_{\rm eff}
&\!=\!&\displaystyle
{1\over 2(2\pi)^2}\sum_{n,m=-\infty}^{\infty}{\!\!\!}'\int_0^\infty ds{s\over \tau^2_n+\tau^2_m}
\Big[{\bar\delta_{m0} \over(s^2-\tau^2_m)}\nonumber\\
&-&{\bar\delta_{n0} \over(s^2+\tau^2_n)}\Big]
e^{-is(m^2_e-i\eta)},
\label{onelooplfini}
\end{eqnarray}
where divergent terms $n\not=0,m=0$, $n=0,m\not=0$ and $n=m=0$ are excluded from the sum, as indicated by
a prime.
The symbol
$\bar\delta_{ij}\equiv 1-\delta_{ij}$ denotes the complimentary Kronecker-$\delta$ which vanishes for $i=j$.
The divergent term with $n=m=0$ is eliminated by the zero-field subtraction in Eq.~(\ref{oneloopl}), while
the divergent terms $n\not=0,m=0$ and $n=0,m\not=0$
\begin{eqnarray}
\Delta{\mathcal L}^{\rm div}_{\rm eff}
&=&
{1\over 2(2\pi)^2}\int_0^\infty {ds\over s}e^{-is(m^2_e-i\eta)}\nonumber\\
&\cdot& 2\left( \sum_{m=1}^{\infty}{1\over\tau^2_m}-\sum_{n=1}^{\infty}
{1\over\tau^2_n}\right),
\label{div}
\end{eqnarray}
are eliminated by the second subtraction in Eq.~(\ref{oneloopl}). This can be seen by performing the sums
\begin{equation}
\sum_{n=1}^\infty\frac{1}{\tau^k_n}=\left(\frac{e\varepsilon}{\pi}\right)^k\zeta(k);\quad
\sum_{n=1}^\infty\frac{1}{\tau^k_m}=\left(\frac{e\beta}{\pi}\right)^k\zeta(k),
\label{sumrule}
\end{equation}
where $\zeta(k)=\sum_n 1/n^k$ is the Riemann function.

The infinitesimal $i\eta$ accompanying the mass term in the $s$-integral (\ref{onelooplfini})
is equivalent to replacing
$e^{-is(m^2_e-i\eta)}$ by $e^{-is(1-i\eta)m^2_e}$. This implies that $s$
is to be integrated slightly below (above) the real axis for $s>0$ ($s<0$). Equivalently one may shift the $\tau_m$ ($-\tau_m$) variables slightly upwards
(downwards) to $\tau_m+i\eta$ ($-\tau_m-i\eta$) in the complex plane.
In order to calculate the finite effective Lagrangian (\ref{onelooplfini}),
the factor $e^{-is(1-i\eta)m^2_e}$ is divided into its sin and cos parts:
\begin{eqnarray}
\Delta{\mathcal L}^{\sin}_{\rm eff}
&=&\displaystyle
{-i\over 4(2\pi)^2}\sum_{n,m=-\infty}^{\infty}{\!\!\!}'\int_{-\infty}^\infty {sds\over \tau^2_n+\tau^2_m}
\Big[{\bar\delta_{m0} \over(s^2-\tau^2_m)}\nonumber\\
&-&{\bar\delta_{n0} \over(s^2+\tau^2_n)}\Big]\sin[s(1-i\eta)m^2_e];
\label{onelooplsin}\\
\Delta{\mathcal L}^{\cos}_{\rm eff}
&=&\displaystyle
{1\over 2(2\pi)^2}\sum_{n,m=-\infty}^{\infty}{\!\!\!}'\int_0^\infty {sds\over \tau^2_n+\tau^2_m}
\Big[{\bar\delta_{m0} \over(s^2-\tau^2_m)}\nonumber\\
&-&{\bar\delta_{n0} \over(s^2+\tau^2_n)}\Big]
\cos[s(1-i\eta)m^2_e].
\label{onelooplcos0}
\end{eqnarray}
The sin part (\ref{onelooplsin}) has an even integrand allowing for
an extension of the $s$-integral over the entire $s$-axis.
The contours of integration can then be closed by infinite semicircles in the half-plane, the integration
receives contributions from poles $\pm \tau_m, \pm i\tau_n$, so that the residue theorem leads to
\begin{eqnarray}
\Delta{\mathcal L}^{\sin}_{\rm eff}\!\!\! &=\!\!\! \displaystyle i{e^2\varepsilon\beta \over 2(2\pi)^2}
\sum^\infty_{n=1}\frac{1}{n}\coth\left({n\pi \beta\over \varepsilon}\right) \exp(-n\pi E_c/\varepsilon)\label{sinE}\\
\!\!\!&\displaystyle\!\!\! -i{e^2\varepsilon\beta \over 2(2\pi)^2}\sum^\infty_{m=1}\frac{1}{m}\coth\left({m\pi \varepsilon\over \beta}\right) \exp(i m\pi E_c/\beta)\label{sinB}.
\end{eqnarray}
The first part (\ref{sinE}) produces the exact non-perturbative Schwinger rate (\ref{probabilityeh})
for pair production.

Shifting $s\rightarrow s-i\eta$, we rewrite the cos part of effective Lagrangian (\ref{onelooplcos0}) as
\begin{eqnarray}
\Delta{\mathcal L}^{\cos}_{\rm eff}
&=&{1\over2(2\pi)^2}
\sum_{n,m=-\infty}^{\infty}{\!\!\!}' \int_0^\infty ds  \frac{\cos(sm^2_e)}{\tau^2_n+\tau^2_m}\nonumber\\
&\cdot&\left({s\bar\delta_{m0} \over s^2-\tau^2_m-i\eta}-{s\bar\delta_{n0} \over s^2+\tau^2_n-i\eta}\right).
\label{onelooplcos}
\end{eqnarray}
In the first term of magnetic part, singularities $s=\tau_m, (m>0)$ and $s=-\tau_m, (m<0)$ appear
in integrating $s$-axis. We decomposes,
\begin{eqnarray}
{s\over s^2-\tau^2_m-i\eta} &= &i\frac{\pi}{2}\delta(s-\tau_m)+i\frac{\pi}{2}\delta(s+\tau_m)\nonumber\\
&+&{\mathcal P}\frac{s}{ s^2-\tau^2_m},\label{h+}
\end{eqnarray}
where ${\mathcal P}$ indicates the principle value under the integral. The integrals over the $\delta$-functions
give
\begin{equation}
\Delta_\delta{\mathcal L}^{\cos}_{\rm eff}=i{e^2\varepsilon\beta \over 2(2\pi)^2}\sum^\infty_{m=1}\frac{1}{m}\coth\left({m\pi \varepsilon\over \beta}\right) \exp(i m\pi E_c/\beta),
\label{hagensin}
\end{equation}
which exactly cancel the second part (\ref{sinB}) of the sin part $\Delta{\mathcal L}^{\sin}_{\rm eff}$.

It remains to find the principal-value integrals in Eq.~(\ref{onelooplcos}),
\begin{eqnarray}
(\Delta{\mathcal L}^{\cos}_{\rm eff})_{\mathcal P}
&=&{1\over2(2\pi)^2}
\sum_{n,m=-\infty}^{\infty}{\!\!\!}' \frac{1}{\tau^2_n+\tau^2_m}{\mathcal P}\int_0^\infty ds
\cos(sm^2_e)\nonumber\\
&\cdot&\left({s\bar\delta_{m0} \over s^2-\tau^2_m}-{s\bar\delta_{n0} \over s^2+\tau^2_n}\right).
\label{onelooplcosp}
\end{eqnarray}
We rewrite the cos function as $\cos(sm^2_e)=(e^{ism^2_e}+e^{-ism^2_e})/2$ and make the rotations of
integration contours by $\pm\pi/2$ respectively,
\begin{eqnarray}
(\Delta{\mathcal L}^{\cos}_{\rm eff})_{\mathcal P}
&=&{1\over2(2\pi)^2}
\sum_{n,m=-\infty}^{\infty}{\!\!\!}' \frac{1}{\tau^2_n+\tau^2_m}{\mathcal P}\int_0^\infty d\tau \nonumber\\
&\cdot&
\left({\bar\delta_{m0}\tau e^{-\tau} \over \tau^2-(i\tau_mm_e^2)^2}
-{\bar\delta_{n0}\tau e^{-\tau} \over \tau^2-(\tau_nm_e^2)^2}\right).
\label{onelooplcosp1}
\end{eqnarray}
Using the formulas (see 3.354, 8.211.1 and 8.211.2 in Ref. \cite{gr1})
\begin{equation}
J(z) \equiv {\mathcal P}\int^\infty_0 ds {s e^{-s}\over s^2-z^2}
= -\frac{1}{2}\Big[e^{-z}{\rm Ei}(z)
+ e^{z}{\rm Ei}(-z)\Big],
\label{J(z)1}
\end{equation}
where ${\rm Ei}(z)$ is the exponential-integral function,
\begin{equation}
{\rm Ei}(z) \equiv {\mathcal P}\int_{-\infty}^z dt \frac{e^t}{t}=\log(-z)+\sum_{k=1}^\infty\frac{z^k}{kk!},
\label{J(z)2}
\end{equation}
we obtain the principal-value integrals (\ref{onelooplcosp1}),
\begin{eqnarray}
(\Delta{\mathcal L}^{\cos}_{\rm eff})_{\mathcal P}  &=&\displaystyle
{1\over2(2\pi)^2}\sum_{n,m=-\infty}^{\infty}{\!\!\!}' {1\over \tau^2_m+\tau^2_n}\nonumber\\
&\cdot&\Big[\bar\delta_{m0}J(i\tau_m m^2_e)-\bar\delta_{n0}J(\tau_n m^2_e)\Big].
\label{pertur}
\end{eqnarray}

With the help of the series and asymptotic representation (see formula 8.215 in Ref. \cite{gr1})
of the exponential-integral function ${\rm Ei}(z)$ for
large $z$, corresponding to weak electromagnetic fields ($\varepsilon\ll 1, \beta\ll 1$),
\begin{equation}
J(z) =-\frac{1}{z^2}-\frac{6}{z^4}-\frac{120}{z^6}-\frac{5040}{z^8}-\frac{362880}{z^{10}}+\cdot\cdot\cdot ,
\label{J(z)3}
\end{equation}
and Eq.~(\ref{pertur}), we find,
\begin{eqnarray}
(\Delta{\mathcal L}^{\cos}_{\rm eff})_{\mathcal P}  &=& \displaystyle
{1\over2(2\pi)^2}\sum_{n,m=-\infty}^{\infty}{\!\!\!}' {1\over \tau^2_m+\tau^2_n}\nonumber\\
\!\!\!&\!\!\!\cdot\!\!\!&\!\!\!\Big\{\bar\delta_{n0}\left[\frac{1}{\tau_n^2m_e^4}+\frac{6}{\tau_n^4m_e^8}+\frac{120}{\tau_n^6m_e^{12}}+\cdot\cdot\cdot\right]\nonumber\\
\!\!\!&\!\!\!+\!\!\!&\!\!\!\displaystyle
\bar\delta_{m0}\left[\frac{1}{\tau_m^2m_e^4}\!-\!\frac{6}{\tau_m^4m_e^8}
\!+\!\frac{120}{\tau_m^6m_e^{12}}\!+\!\cdot\cdot\cdot\!\right]\!\Big\}.
\label{perexpansion}
\end{eqnarray}
Applying the summation formulas (\ref{sumrule}),
the weak-field expansion (\ref{perexpansion}) is seen to agree with
the Heisenberg and Euler effective Lagrangian \cite{euler},
\begin{eqnarray}
(\Delta{\mathcal L}_{\rm eff})_{\mathcal P} &=&\displaystyle
\frac{2 \alpha ^2}{45 m_e^4}\left\{
({\bf E}^2\!-\!{\bf B}^2)^2+7 ({\bf E}\cdot {\bf B})^2 \right\}\nonumber\\
&+&\displaystyle \frac{64\pi \alpha ^3}{315 m_e^8}\left\{
2({\bf E}^2\!-\!{\bf B}^2)^3+13({\bf E}^2\!-\!{\bf B}^2)
 ({\bf E}\cdot {\bf B})^2 \right\}\nonumber\\
 &+&\cdot\cdot\cdot ,
\label{Kleinert1}
\end{eqnarray}
which is expressed in terms of a powers series of week
electromagnetic fields up to $O(\alpha^3)$. The expansion coefficients
of the terms of order $n$
have the general form $m_e^4/(E_c)^n$. As long as the fields
are much smaller than $E_c$, the expansion
converges.

On the other hand, using the series and asymptotic
representation (formulas 8.214.1 and 8.214.2 in Ref. \cite{gr1})
of the exponential-integral function ${\rm Ei}(z)$ for
small $z\ll 1$, corresponding to strong electromagnetic fields ($\varepsilon\gg 1, \beta\gg 1$),
\begin{equation}
J(z) =-\frac{1}{2}\Big[e^z\ln(z)+e^{-z}\ln(-z)\Big]+\gamma +{\mathcal O}(z),
\label{smallz}
\end{equation}
with $\gamma=0.577216$ being the Euler-Mascheroni constant,
we obtain the leading terms in the strong-field expansion of Eq.~(\ref{pertur}),
\begin{eqnarray}
(\Delta{\mathcal L}^{\rm cos}_{\rm eff})_{\mathcal P}
&=&{1\over2(2\pi)^2}\sum_{n,m=-\infty}^{\infty '} {1\over \tau^2_m+\tau^2_n}\nonumber\\
\!\!\!&\!\!\!\cdot\!\!\!&\!\!\!\Big[\bar\delta_{n0}\ln(\tau_n m^2_e)\!-\!\bar\delta_{m0}\ln(\tau_m m^2_e)\Big]\!+\!\cdot\cdot\cdot .
\label{strongexp}
\end{eqnarray}
In the case of vanishing magnetic field ${\bf B}=0$ and $m=0$ in Eq.~(\ref{strongexp}), we have
\begin{eqnarray}
(\Delta{\mathcal L}^{\rm cos}_{\rm eff})_{\mathcal P}  &=&{1\over2(2\pi)^2}\sum_{n=1}^{\infty} {1\over \tau^2_n}
\ln(\tau_n m^2_e)+\cdot\cdot\cdot\nonumber\\
 &=&{e^2E^2\over8\pi^4}\sum_{n=1}^{\infty} {1\over n^2}
\ln\left(\frac{n\pi E_c}{E}\right)+\cdot\cdot\cdot ,
\label{strongexpe}
\end{eqnarray}
for a strong electric field $\bf E$. In the case of vanishing electric field ${\bf E}=0$ and $n=0$ in Eq.~(\ref{strongexp}),
we obtain for a strong magnetic field $\bf B$,
\begin{eqnarray}
(\Delta{\mathcal L}^{\rm cos}_{\rm eff})_{\mathcal P}  &=&-{1\over2(2\pi)^2}\sum_{m=1}^{\infty} {1\over \tau^2_m}
\ln(\tau_m m^2_e)+\cdot\cdot\cdot\nonumber\\
& =&-{e^2B^2\over8\pi^4}\sum_{m=1}^{\infty} {1\over m^2}
\ln\left(\frac{m\pi E_c}{B}\right)+\cdot\cdot\cdot ,
\label{strongexpb}
\end{eqnarray}
a result first considered by Weisskopf \cite{weisskopf0}.

\subsubsection{Concluding remarks}

We have presented in Eqs.~(\ref{sinE},\ref{sinB},\ref{hagensin},\ref{pertur})
closed form result for the one-loop effective Lagrangian $\Delta{\mathcal L}_{\rm eff}$ (\ref{oneloopl})
for arbitrary strength of constant electromagnetic fields.
The results will receive fluctuation corrections
from higher loop diagrams. These carry one or more factors $ \alpha $, $ \alpha ^2$, \dots~ and are thus
suppressed by factors $1/137$.
Thus results are valid for all field strenghts
with an error no larger than roughly  1\%.
If we include, for example, the two-loop correction, the first term in the Heisenberg and
Euler effective Lagrangian (\ref{Kleinert1}) becomes \cite{hagen}
\begin{eqnarray}
(\Delta{\mathcal L}_{\rm eff})_{\mathcal P} &=&
\frac{2 \alpha ^2}{45 m_e^4}\Big\{
\left(1+\frac{40 \alpha }{9\pi}\right)({\bf E}^2\!-\!{\bf B}^2)^2\nonumber\\
&+&7
\left(1+\frac{1315 \alpha }{252\pi} \right)({\bf E}\cdot {\bf B})^2 \Big\}.
\label{Kleinert2}
\end{eqnarray}
Readers can consult the recent review article \cite{dunne}, where
one finds discussions and computations of the effective Lagrangian at tow-loop levels,
in homogeneous and inhomogeneous fields.

The interaction of an external electromagnetic field with the QED vacuum leads
to the appearance of nonlinear real and imaginary parts in the effective Lagrangian
$\Delta{\mathcal L}_{\rm eff}$ of the electromagnetic field.
Electron and positron pair production is only one aspect of the phenomena connecting to these nonlinearities.
As we remarked before the rate $\Gamma/V$ of pair production (\ref{probability1}) is exponentially small
when $E \lesssim E_c$.

Unfortunately, it seems inconceivable to produce a macroscopic static field with electric field strengths of
the order of the critical value in any ground laboratory to directly observe the
Sauter-Euler-Heisenberg-Schwinger process of electron-positron pair production in vacuum.
In Ref.~\cite{rvx2005}, we discuss some ideas: (i) heavy-ion collisions \cite{grc98},
(ii) the focus of coherent laser beams \cite{r} and (iii) electron beam-laser collisions \cite{dla},
to experimentally create a transient electric field $E\lesssim E_c$ in earth-bound laboratories,
whose lifetime is expected to be long enough (larger than $\hbar/m_ec^2$) for the pair production
process to take place. In Ref.~\cite{rvx2005}, we also discuss (i) the idea that the supercritical conditions can be reached
on the time scale of gravitational collapse ($GM/c^2$, $10^{-2}{\rm sec.}$
for $M=10M_\odot$) around an astrophysical object \cite{RVX03}
and (ii) the application of Sauter-Euler-Heisenberg-Schwinger process
to the black hole physics in connection with the observed astrophysical phenomenon of gamma ray bursts \cite{Brasile}.

Acknowledgement: we thank Prof. H. Kleinert for many discussions and cooperations on this article.

%\newpage

%\newpage

%\begin{figure}
%\vspace{5cm}
%\caption{$\delta \alpha /\alpha$ (times $10^{-5}$) vs fractional
%look-back time to big bang, predicted by present work as compared
%with the data by Webb et al \cite{Web01} (explanation in the text).}
%\label{fig1}
%\end{figure}

%\bibitem{rw}
%T.~Regge and J.A.~Wheeler, Phys.~ Rev.~{\bf 108}, 1063 (1957).

\end{document}